\documentclass[aps,prd,preprint,superscriptaddress,amsmath,amssymb,showpacs]{revtex4-1}
\usepackage{dcolumn}
\usepackage{graphicx}
\usepackage{float}
\usepackage{physics}
\usepackage[colorlinks=true,allcolors=blue]{hyperref}
\begin{document}
	\title{Rotation effect on the spectral function of heavy vector mesons \\
		in holographic QCD
	}
	
	\author{Xiao-Long Wang}
	\affiliation{College of Science, China Three Gorges University, Yichang 443002, China}
	
	\author{Sheng-Qin Feng}
	\email{Corresponding author: fengsq@ctgu.edu.cn}
	\affiliation{College of Science, China Three Gorges University, Yichang 443002, China}
	\affiliation{Center for Astronomy and Space Sciences and Institute of Modern Physics, China Three Gorges University, Yichang 443002, China}
	\affiliation{Key Laboratory of Quark and Lepton Physics (MOE) and Institute of Particle Physics,\\
		Central China Normal University, Wuhan 430079, China}
	
	\date{\today}

	\begin{abstract}
		Abstract:  Exploring heavy vector mesons of the  $ J / \psi$ and $ \Upsilon ( 1 S )$ is crucial for understanding the quark gluon plasma (QGP) formed in heavy ion collisions. The influences of rotational effect on the properties of the $ J / \psi$ and the $ \Upsilon ( 1 S )$ are investigated by incorporating rotation medium into the holographic QCD. It is found that temperature, chemical potential, and rotational radius effects enhance the dissociation process of the $ J / \psi$ and the $ \Upsilon ( 1 S )$ states within the medium. This rotation-induced effect is more significant for heavy vector mesons in the transverse direction than that of the longitudinal direction.
The first holographic study on the influence of the radius of a homogeneous rotating system on the vector meson spectrum is proposed. It is found that increasing in rotation radius promotes the dissociation of vector mesons of the $ J / \psi$ and $ \Upsilon ( 1 S )$. We also find that the dissociation perpendicular to the direction of rotational angular velocity is more significant than that parallel to it at large rational radius.
	\end{abstract}
	
	\maketitle
	
	\section{Introduction}\label{sec:01_intro}
	
As is well known, relativistic heavy ion collisions produce a high-temperature, high-density medium, known as quark-gluon plasma (QGP) \cite{RN1,RN2,RN3}. In this QGP state, gluons and light quarks are in a deconfinement phase, exhibiting perfect fluid behavior that is approximately thermalized, providing an important tool for exploring quantum chromodynamics (QCD). These QGP matters are mutually correlated through strong interactions, but they are not confined within hadrons. They have an extremely short-lived $ ( \sim5-10 fm/c$), and direct observation of the plasma is impossible. Heavy Mesons composed of $c$ (charm) or $b$ (bottom) quarks are more likely to survive long time in the hot QGP medium compared to mesons composed of lighter quarks : $u$ (up), $d$  (down) and $s$ (strange). It is believed that heavy mesons can serve as probes to indirectly gather information about the Quark-Gluon Plasma (QGP) \cite{RN4,RN5,RN6}.

Due to the influence of temperature, chemical potential, and other relevant properties on the properties of QCD media, these heavy vector mesons undergo dissociation in high-temperature media. Therefore, understanding the thermal behavior of heavy vector mesons, especially the relationship between their dissociation degree and medium dynamics, is crucial for us to have a deeper understanding of the properties of QCD media.

In relativistic heavy ion collisions, in addition to the presence of strong magnetic fields \cite{RN7,RN8,RN9,RN10,RN11}, the plasma also exhibits significant rotational vortex characteristics. Some studies have introduced rotation into holographic research \cite{RN12,RN13,RN14,RN15,RN16,RN17} and effective field theory \cite{RN18,RN19,RN20,RN21,RN22,RN23}. The phases of matter and the properties of associated particles become very complex under rotation, which has recently attracted great interest \cite{RN12,RN18}. These studies are closely related to the properties of strong interactions in quantum chromodynamics (QCD). For example, astrophysical objects such as neutron stars composed of dense QCD matter can rotate rapidly \cite{RN24,RN25}. The QCD medium produced by non-central nucleus-nucleus collisions also carries a large amount of rotational angular momentum, with an order of magnitude of $10^{4}-10^{5}\hbar$, and local rotational angular velocity can reach the order of magnitude of 0.01-0.1 GeV \cite{RN26,RN27,RN28,RN29,RN30,RN31,RN32}. On the other hand, significant progress has also been made in simulating rotating QCD matter using lattice gauge theory \cite{RN33}.

The gauge/gravity duality has become an important research tool for exploring QCD phase transitions \cite{RN12,RN34,RN35,RN36} and transport properties \cite{RN37,RN38,RN39}. Recently, holographic methods have been widely used to study the properties of heavy flavor spectral functions under strong magnetic fields and medium rotation conditions \cite{RN40,RN41,RN42,RN43,RN44,RN45,RN46,RN47,RN48,RN49,RN50,RN51,RN52}. The gauge/gravity duality has not only become a valuable tool for exploring QCD phase transitions and transport properties, but also been widely used to explore the properties of heavy meson spectral functions.
\vspace{-2mm}

In this article, we will use the holographic model to study the rotating QCD matter, where the rotating AdS black hole is the holographic dual of the plasma. The heavy vector meson spectral function of the rotating QCD system by introducing a metric with rotating cylindrical coordinates is studied. Although there are many non-perturbative methods to explore the heavy vector meson spectral function~\cite{RN53,RN54}, the main purpose of this article is to study the influence of plasma on the heavy vector meson dissociation under rotation and finite chemical potential. We focus on studying the dependence of the dissociation process of heavy vector mesons $ J / \psi$ and $ \Upsilon ( 1 S )$  on temperature, chemical potential, and rotation radius.
\vspace{-2mm}

Refenence \cite{RN12} first used holography to study the influence of the radius of a uniform rotating system on the QCD phase diagram. It is found that when discussing the rotating system of QCD medium, the phase transition characteristics strongly depend on the finite size of the rotating system. Due to the cylindrical symmetry of the rotating system, the rotation radius $\mathit{l}$ has become an important characteristic of the rotating system. We will use holography to mainly discuss the influence of the radius of a uniform rotating system on the heavy vector meson spectrum, and understand how the finite size of the rotating system affects the characteristics of the QCD system spectrum function in this article.

The structure of the paper is as follows: in Sec. II , we develop the holographic model in plasma and extend it to finite temperature and finite chemical potential.  A holographic model of  the spectral function of heavy vector mesons in a rotating plasma is introduced In Sec. III. In Sec. IV , we present and discuss the calculated results of spectral functions of the heavy vector mesons. Finally, in Section V, we provide summary and conclusions.
		
	\section{Holographic QCD theory under rotational background}\label{sec:02 setup}
	
The generalized version of the soft wall model \cite{RN40} is introduced, where the vector meson is described by a vector field $ V _ { m } = ( V _ { \mu } , V _ { z } ) ( \mu = 0 , 1 , 2 , 3 )$, which is dual to the current $ J ^ { \mu } = \psi \gamma ^ { \mu } \psi$ of the gauge theory. The standard Maxwell action is
	\begin{equation}\label{eq:01}
		S = - \int d ^ { 4 } x d z \frac { Q } { 4 } F _ { m n } F ^ { m n } ,
	\end{equation}
	where $Q = \frac { \sqrt { - g } } { h ( \phi ) g _ { 5 } ^ { 2 } } $, ~~$ h ( \phi ) = e ^ { \phi ( z ) }, F _ { m n } = \partial _ { m } V _ { n } - \partial _ { n } V _ { m }$, and $\phi ( z ) $ the dilaton background.

The metric is given by in the background of a charged black holes:
	\begin{equation}\label{eq:02}
		d s ^ { 2 } = \frac { R ^ { 2 } } { z ^ { 2 } } ( - f ( z ) d t ^ { 2 } + \frac { d t ^ { 2 } } { f ( z ) } + d \vec{x}  \cdot d \vec{x}  ) ,
	\end{equation}
with
\begin{equation}\label{eq:03}
	f ( z ) = 1 - \frac { z ^ { 4 } } { z _ { h } ^ { 4 } } - q ^ { 2 } z _ { h } ^ { 2 } z ^ { 4 } + q ^ { 2 } z ^ { 6 } ,
\end{equation}
and
\begin{equation}\label{eq:04}
	\phi ( z ) = k ^ { 2 } z ^ { 2 } + M z + \tanh ( \frac { 1 } { M z } - \frac { k } { \sqrt { \Gamma } } ) ,
\end{equation}
where $k$ represents the quark mass, $M$ denotes a large mass and $\Gamma$ is the string tension of the quark pair associated with the non-hadronic decays of heavy quarkonium. There is a matrix element $ <0 |J_{\mu}(0)|X(1S)> = \in_{\mu}f_{n} m_{n}$ (where represents the heavy mesons,  $ <0|$ is the hadronic vacuum and $J_{\mu}(0)$ is the hadronic current and $f _ { n }$ is the decay constant). By fitting the mass spectrum, one determines the best values \cite{RN43}  for the three energy parameters in the scalar field for charmonium and bottomonium as
\begin{equation}\label{eq:05}
	k _ { c } = 1 . 2 G e V , \sqrt { \Gamma _ { c } } = 0 . 5 5 G e V , M _ { c } = 2 . 2 G e V ,
\end{equation}
\begin{equation}\label{eq:06}
	 k _ { b } = 2 . 4 5 G e V , \sqrt {  \Gamma _ { b } } = 1 . 5 5 G e V , M _ { b } = 6 . 2 G e V .
\end{equation}
	
	The horizon position $z _ { h }$  of the black hole is obtained by the condition $ f ( z _ { h } ) = 0 $. The temperature of the black hole is determined by the requirement of the Euclidean version of the metric having no conical singularity at the horizon. It satisfies the following relation:
\begin{equation}\label{eq:07}
T = \frac { | f ^ { \prime } ( z ) | _ { ( z = z _ { h } ) } } { 4 \pi } = \frac { 1 } { \pi z _ { h } } - \frac { q ^ { 2 } z _ { h } ^ { 5 } } { 2 \pi } ,
\end{equation} 	
where the parameter $q$ is proportional to the black hole charge, which is related to the quark chemical potential $ \mu$. In the dual supergravity description, the field $V _ { 0 } $ acts as a source for the correlator’s operators. Therefore, a specific solution of the vector field $V _ { M } $ was considered, where the non-zero component is given by: $ V _ { 0 } = A _ { 0 } ( z ) ( V _ { z } = 0 , V _ { i } = 0 )$. Assuming that the relationship between $ q $ and $ \mu$ is the same as in the background-free case, i.e.,~$ \phi ( z ) = 0 ,$ the solution for the time component of the vector field is given by:$ A _ { 0 } ( z ) = c - q z ^ { 2 }$, where $c$  is a constant. By requiring  $ A _ { 0 } (0 ) = \mu $ and $ A _ { 0 } ( z _ { h } ) = 0 $, one obtain:
\begin{equation}\label{eq:08}
 \mu = q z_{h}^{2}  .
\end{equation}

In the early stages of non-central heavy ion collisions, the produced quarks carry a large initial orbital angular momentum  $ J \propto b \sqrt { S _ { N N } } , $  where $b$ is the impact parameter and $\sqrt { S _ { N N } } $ is the center-of-mass energy of the nucleon-nucleon collision. Although most of the angular momentum is carried away by the so-called "spectators" during the initial collision stage, a significant portion of the angular momentum remains in the resulting quark-gluon plasma.

Basing on the study in \cite{RN14,RN15,RN54,RN55,RN56} , we extend the holographic QCD model to include the case of a rotating black hole with a planar horizon. The general form of the metric of the static coordinate system is given by:
\begin{equation}\label{eq:09}
d s ^ { 2 } = - g _ { tt } d t ^ { 2 } + g _ { z z } d z ^ { 2 } + g _ { x _ { 1 } x _ { 1 } } d x _ { 1 } ^ { 2 } + g _ { x _ { 2 } x _ { 2 } } d x _ { 2 } ^ { 2 } + g _ { x _ { 3 } x _ { 3 } } d x _ { 3 } ^ { 2 } .
\end{equation}

Now, in order to analyze the case of a rotating plasma with uniform angular velocity, we introduce an AdS$_5$ space with cylindrical symmetry and describe its structure by rewriting the metric as
\begin{equation}\label{eq:10}
 d s ^ { 2 } = - g _ { tt } d t ^ { 2 } + g _ { zz } d z ^ { 2 } + g _ { \theta \theta }\mathit{l}^{2}  d \theta ^ { 2 } + g _ { x _ { 2 } x _ { 2 } } d x _ { 2 } ^ { 2 } + g _ { x _ { 3 } x _ { 3 } } d x _ { 3 } ^ { 2 } .
\end{equation}

Then the angular momentum will be turned on in the angular coordinate $\theta$ through the standard Lorentz transformation. As in Refs. \cite{RN42,RN47}, we introduce rotation through a Lorentz-like coordinate transformation as
\begin{equation}\label{eq:11}
 t \rightarrow \gamma ( t + \Omega l ^ { 2 } \theta ) ,
\end{equation}
and
\begin{equation}\label{eq:12}
 \theta \rightarrow \gamma ( \Omega t + \theta ) ,
\end{equation}
where $ \gamma = \frac { 1 } { \sqrt { 1 - \Omega ^ { 2 } l ^ { 2 } } }$ is the usual Lorentz factor, $\Omega$ is the angular velocity of rotation, and $l$ is the radius of the rotating axis. With the introduction of rotation, the original metric of Eq. (9) becomes the following form
	\begin{equation}\label{eq:13}
	\begin{aligned}
	 d s ^ { 2 } =  & ( g _ { \theta \theta } \Omega ^ { 2 } l ^ { 2 } - g _ { t t } ) \gamma ^ { 2 } d t ^ { 2 } + 2 \gamma ^ { 2 } \Omega l ^ { 2 } ( g _ { \theta \theta } - g _ { t t } ) d t d \theta\\
		&\ + \gamma ^ { 2 } ( g _ { \theta \theta } - \Omega ^ { 2 } l ^ { 2 } g _ { tt } ) l ^ { 2 } d \theta ^ { 2 } + g _ { z z } d z ^ { 2 } + g _ { x _ { 2 } x _ { 2 } } d x _ { 2 } ^ { 2 } + g _ { x _ { 3 } x _ { 3 } } d x _ { 3 } ^ { 2 } .
	\end{aligned}
\end{equation}

The Hawking temperature and chemical potential of a rotating black hole are given by
\begin{equation}\label{eq:14}
 T = \big( \frac { 1 } { \pi z _ { h } } - \frac { q ^ { 2 } z _ { h } ^ { 5 } } { 2 \pi } \big ) \sqrt { 1 - \Omega ^ { 2 } l ^ { 2 } } ,
\end{equation}
and
\begin{equation}\label{eq:15}
 \mu = q z_{h}^{2} \sqrt { 1 - \Omega ^ { 2 } l ^ { 2 } } .	
\end{equation}	
	\section{The spectral functions of heavy vector mesons}\label{sec:03 setup}
	
	By making a variable substitution $ l \theta = x _ { 1 }$ with Eq. (10), one can obtain a new metric representation along the $x _ { 1 }$ direction
\begin{equation}\label{eq:16}
d s ^ { 2 } = - G _ { t t } d t ^ { 2 } + G _ { x _ { 1 } x _ { 1 } } d x _ { 1 } ^ { 2 } + G _ {t x _ { 1 } } d t d x _ { 1 } + G _ { x _ { 1 } t } d x _ { 1 } d t + G _ { x _ { 2 } x _ { 2 } } d t _ { 2 } ^ { 2 } + G _ { x _ { 3 } x _ { 3 } } d t _ { 3 } ^ { 2 } + G _ { zz } d z ^ { 2 } .
\end{equation}

The spectral function of the charmonium and bottomonium can be calculated using the membrane paradigm \cite{RN49}. The equations of motion obtained from Eq. (1) are as follows:
\begin{equation}\label{eq:17}
 \partial _ { m } ( Q F ^ { m n } ) = \partial _ { z } ( Q F ^ { z n } ) + \partial _ { \mu } ( Q F ^ { \mu m } ) ,
\end{equation}
where $ F ^ { m n } = g ^ { m \alpha } g ^ { n \beta } F _ { \alpha \beta },  n = ( 0 , 1 , 2 , 3 , 4 )$ and $ \mu = ( 0 , 1 , 2 , 3 )$. For the $z$-direction, the conjugate momentum of the gauge field $ A ^ { \mu }$ is given by the following equation:
\begin{equation}\label{eq:18}
 j ^ { \mu } = - Q F ^ { z \mu } .
\end{equation}

Now we consider the plane wave solution of vector field $ A ^ { \mu }$ propagates in the direction of  $ x _ { 1 }$, and divide the equation of motion into a longitudinal channel involving fluctuations along $ ( t , x _ { 1 } )$ and a transverse channel involving fluctuations along the spatial directions $ ( x _ { 2 } , x _ { 3 } ) .$  Combining Eq. (18) with Eq. (17), one can obtain the dynamical equation for the longitudinal case as:
\begin{equation}\label{eq:19}
	 - \partial _ { z } j ^ {t } - \frac { \sqrt { - g } } { h ( \phi ) } \big ( g ^ { x _ { 1 } x _ { 1 } } g ^ { tt } + g ^ { x _ { 1 } t } g ^ { t x _ { 1 } } \big ) \partial _ { x _ { 1 } } F _ { x _ { 1 } t } = 0 ,
\end{equation}
\begin{equation}\label{eq:20}
	 - \partial _ { z } j ^ { x _ { 1 } } + \frac { \sqrt { - g } } { h ( \phi ) } \big ( g ^ { tt } g ^ { x _ { 1 } x _ { 1 } } + g ^ {t x _ { 1 } } g ^ { x _ { 1 } t } \big ) \partial _ { t } F _ { x _ { 1 } t } = 0 ,
\end{equation}
\begin{equation}\label{eq:21}
 \partial _ { x _ { 1 } } j ^ { x _ { 1 } } + \partial _ { t } j ^ { t } = 0 .
\end{equation}
Using Bianchi's identity, one can obtain:
\begin{equation}\label{eq:22}
 \partial _ { z } F _ { x _ { 1 } t } - \frac { g _ { zz } h ( \phi ) } { \sqrt { - g } } \partial _ { t } \left[ g _ { x _ { 1 } x _ { 1 } } j ^ { x _ { 1 } } + g _ { x _ { 1 } t } j ^ { t } \right] - \frac { g _ { zz } h ( \phi ) } { \sqrt { - g } } \partial _ { x _ { 1 } } \left[ g _ { tt } j ^ { t } - g _ { tx _ { 1 } } j ^ { x _ { 1 } } \right] = 0 .
\end{equation}

Now, it is possible to define the conductivity of the longitudinal channel with respect to $z$
\begin{equation}\label{eq:23}
 \sigma _ { L } ( \omega , z ) = \frac { j ^ { x _ { 1 } } ( \omega , z ) } { F _ { x _ { 1 }t } ( \omega , z ) } ,
\end{equation}
and its derivative is defined as
\begin{equation}\label{eq:24}
 \partial _ { z } \sigma _ { L } ( \omega , z ) = \frac { \partial _ { z } j ^ { x _ { 1 } } } { F _ { x _ { 1 } t } } - \frac { j ^ { x _ { 1 } } } { F _ { x _ { 1 } t } ^ { 2 } } \partial _ { z } F _ { x _ { 1 } t } .
\end{equation}

The Kubo formula $ \sigma ( \omega ) = i G _ { R } ( \omega ) / \omega$ displays the five-dimensional conductivity at the boundary, which allows us to define
\begin{equation}\label{eq:25}
 \sigma _ { L } ( w ) = \frac { - G _ { R } ^ { L } ( w ) } { i w } .
\end{equation}

Notice that the real part is $\textrm{Re} \sigma ( \omega ) = \rho ( \omega ) / \omega ,$ where $ \rho ( \omega ) \equiv - \textrm{Im} G _ { R } ( \omega )$ is the spectral function. To obtain the flow equation Eq. (24), one assumes that $ A _ { \mu } = A _ { n } ( p , z ) e ^ { - i \omega t + i p x _ { 1 } } ,$ where $A _ { n } ( p , z )$  represents the quasi-normal mode. Therefore, one obtains $ \partial _ { t } F _ { x _ { 1 } t } = - i \omega F _ { x _ { 1 } t } $ , $ \partial _ { t } j ^ { x _ { 1 } } = - i \omega j ^ { x _ { 1 } } .$  By utilizing Eqs. (20), (21), (22) and taking the momentum limit, the Eq. (24) can be calculated as

\begin{equation}\label{eq:26}
 \partial _ { z } \sigma _ { L } ( \omega , z ) = i \omega \Delta _ { L } ( \sigma _ { L } ( \omega , z ) ^ { 2 } - ( \Sigma  _ { L } ) ^ { 2 } ) ,
\end{equation}
where
\begin{equation}\label{eq:27}
 \Delta _ { L } = \frac { e ^ { \phi  ( z ) } z ( l^ { 2 } \Omega ^ { 2 } f ( z ) - 1 ) } { ( l ^ { 2 } \Omega ^ { 2 } - 1 ) f ( z ) } ,
\end{equation}
and
\begin{equation}\label{eq:28}
\Sigma  _ { L } = \frac { e ^ { - \phi ( z ) } } { z } \sqrt { \frac { l ^ { 2 } \Omega ^ { 2 } - 1 } { l ^ { 2 } \Omega ^ { 2 } f ( z ) - 1 } }.
\end{equation}

The dynamical equation for the transverse channel is given by
	\begin{equation}\label{eq:29}
	\begin{aligned}
		\partial _ { z } j ^ { x _ { 2 } } & - \frac { \sqrt { - g } } { h ( \phi ) } \Big [ g ^ {t x _ { 1 } } g ^ { x _ { 2 } x _ { 2 } } \partial _ { t } F _ { x _ { 1 } x _ { 2 } } - g ^ { tt } g ^ { x _ { 2 } x _ { 2 } } \partial _ { t } F _ { tx _ { 2 } } \\
		& + g ^ { x _ { 1 } t } g ^ { x _ { 2 } x _ { 2 } } \partial _ { x _ { 1 } } F _ { t x _ { 2 } } + g ^ { x _ { 1 } x _ { 1 } } g ^ { x _ { 2 } x _ { 2 } } \partial _ { x _ { 1 } } F _ { x _ { 1 } x _ { 2 } } \Big ] = 0
	\end{aligned}
\end{equation}
\begin{equation}\label{eq:30}
 \frac { h ( \phi ) g _ { x _ { 2 } x _ { 2 } } g_ { zz }  } { \sqrt { - g } } \partial _ { t } j ^ { x _ { 2 } } + \partial _ { z } F _ { t x _ { 2 } } = 0 ,
\end{equation}
\begin{equation}\label{eq:31}
 \partial _ { x _ { 1 } } F _ { tx _ { 2 } } + \partial _ { t } F _ { x _ { 2 } x _ { 1 } } = 0 .
\end{equation}

The transverse "conductivity" and its derivative are defined as
\begin{equation}\label{eq:32}
 \sigma _ { T } ( \omega , z ) = \frac { j ^ { x _ { 2 } } ( \omega , \vec{p} ,z ) } { F _ { x _ { 2 } t } ( \omega , \vec{p}  , z ) } ,
\end{equation}
\begin{equation}\label{eq:33}
 \partial _ { z } \sigma _ { T } ( \omega , z ) = \frac { \partial _ { z } j ^ { x _ { 2 } } } { F _ { x _ { 2 } t } } - \frac { j ^ { x _ { 2 } } } { F _ { x _ { 2 } t } ^ { 2 } } \partial _ { z } F _ { x _ { 2 } t } .
\end{equation}

Similarly, we have $ \partial _ { t } F _ { x _ { 1 } x _ { 2 } } = - i \omega F _ { x _ { 1 } x _ { 2 } },  \partial _ { t } F _ { t x _ { 2 } } = - i \omega F _ { t x _ { 2 } } , \partial _ { t } j ^ { x _ { 2 } } = - i \omega j ^ { x _ { 2 } } .$ Therefore, the transverse flow of Eq. (33) can be written as:
\begin{equation}\label{eq:34}
 \partial _ { z } \sigma _ { T } ( \omega , z ) = i \omega \Delta _ { T } \big( \sigma _ { T } ( \omega , z ) ^ { 2 } - ( \Sigma  _ { T } ) ^ { 2 } \big) ,
\end{equation}
where
\begin{equation}\label{eq:35}
 \Delta _ { T } = \frac { e ^ { \phi  ( z ) } z } { f ( z ) } ,
\end{equation}
and
\begin{equation}\label{eq:36}
\Sigma  _ { T } = \frac { e ^ { - \phi ( z ) } } { z } \sqrt { \frac { l ^ { 2 } \Omega ^ { 2 } f ( z ) - 1 } { l ^ { 2 } \Omega ^ { 2 } - 1 } } .
\end{equation}

By requiring the regularity of the horizon, one can obtain the initial conditions for solving the equation $\partial _ { z } \sigma _ {L (T) } =0$. The flow Eqs. (26) and (34) have the same form under a rotation of $ \Omega $ = 0  GeV. The spectral function can be defined by the retarded Green's function as
\begin{equation}\label{eq:37}
\rho ( \omega ) \equiv - \textrm{Im} G _ { R } ( \omega ) = \omega \textrm{Re} \sigma ( \omega , 0 ) .
\end{equation}

	\section{Numerical results of spectral functions in a rotating plasma}\label{sec:04 setup}
In this paper, we describe the spectral functions of heavy vector mesons $( J / \psi$ and $ \Upsilon ( 1 S ))$ of eq. (37) by computing Eq.s (26) and (34) and utilizing the boundary conditions described in the previous section. The spectral functions of heavy vector mesons $( J / \psi$ and $ \Upsilon ( 1 S ))$ with rotation are calculated by using the membrane paradigm \cite{RN57} .
\subsection{The Result For $J / \psi$}
	\begin{figure}[H]
	\centering
	\includegraphics[width=0.7\textwidth]{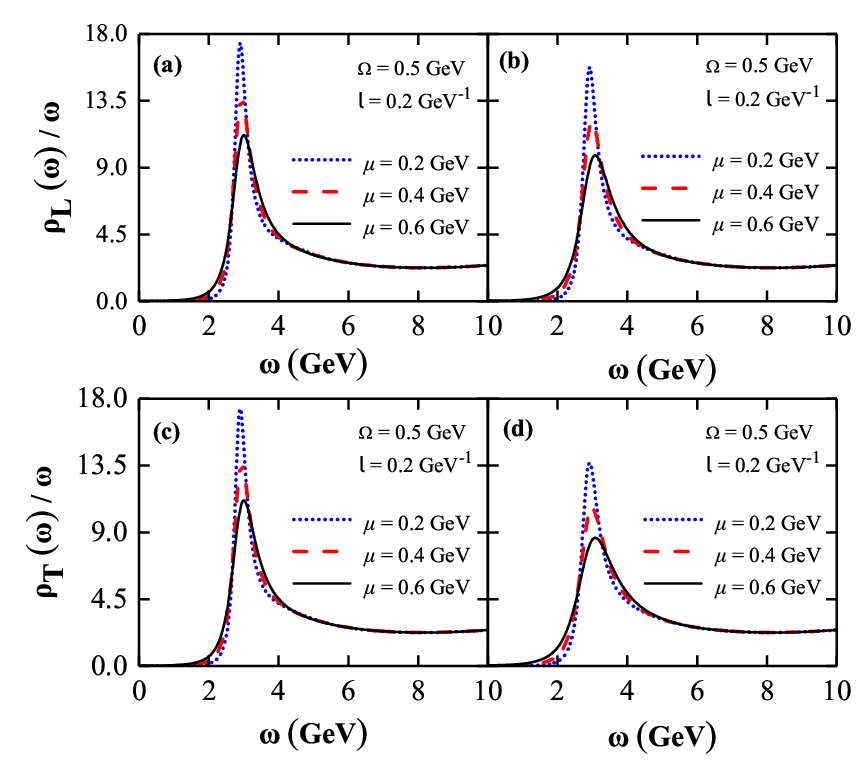}
	\caption{\label{fig1} Spectral functions of $ J / \psi$ at $T$ = 0.2 GeV , under different chemical potentials $\mu $ and rotational radius $l$ . Figs. 1(a) and 1(b) are parallel to the rotational angular velocity, while Figs. 1 (c) and 1(d) are perpendicular to the rotational angular velocity.}
\end{figure}

In holographic rotating systems, the spectral function and the effective mass should also depend on finite dimensions. Due to cylindrical symmetry, these quantities depend on the radius $l$ mentioned in Sect. II . It would be interesting to explore how various thermodynamic quantities of strongly interacting rotating matter vary with the radius of the rotating system. The nature of the rotating system as a function of the radius may be relevant to future experimental observations. In addition, the radius is expected to significantly change the angular momentum and moment of inertia. Therefore, it becomes important to investigate how the spectral function and effective mass of QCD matter under rotation depend on the rotational radius of the system. Due to the limitation of the speed of light, it naturally leads to the restriction $ \Omega l \leq 1 $.

Figure 1 presents the spectral functions of the $ J / \psi$ vector mesons at fixed rotational angular velocity $\Omega$, with different chemical potentials $\mu$ and rotational radius $l$ . This clearly demonstrates the impacts of density effects and rotational radius effects on the dissociation process. There is a distinct peak corresponding to the $ J / \psi$ meson state. The bell-shaped curve represents the resonance state, and the peak position corresponds to the resonance mass of the effective mass. Figures 1(a) and 1(b) exhibit the spectral function parallel to the rotational angular velocity, while Figs. 1(c) and 1(d) exhibit the spectral function perpendicular to the rotational angular velocity. We can see that the resonance peaks in Figs. 1(c) and 1(d) are slightly lower than those in Figs. 1(a) and 1(b), especially at large chemical potentials and large rotational radii, which means that the degree of dissociation perpendicular to the rotational angular velocity direction is greater than that parallel to the rotational angular velocity direction. As both the chemical potential $\mu$ and the rotational radius $l$ increase, we observe that the peak decreases, broadens, and slightly shifts to the right. The study indicates that as the chemical potential and rotational radius increase, they promote the dissociation of bound states and increase the decay width and effective mass.
	\begin{figure}[H]
	\centering
	\includegraphics[width=0.75\textwidth]{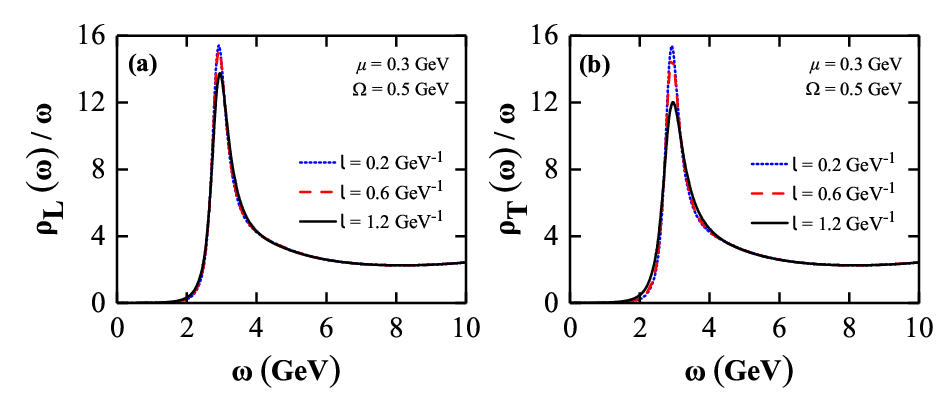}
	\caption{\label{fig2} Spectral function of $ J / \psi$ for different rotation radii with parallel and
		perpendicular to the rotational angular velocity at $\mu$ = 0.3 GeV , $\Omega$ = 0.5 GeV and $T$ = 0.2 GeV .}
\end{figure}

Figure 2 shows the spectral function of the $ J / \psi$ at different rotation radii. The left figure represents the longitudinal spectral function parallel to the rotational angular velocity, and the right figure represents the transverse spectral function perpendicular to the rotational angular velocity. The resonance peak of $ J / \psi$ decreases with increasing rotation radius for  $T$ = 0.2 GeV , $\mu$ = 0.3 GeV and $\Omega$ = 0.5 GeV. In addition, the dissociation effect in the transverse direction is more pronounced than in the longitudinal direction, and this effect becomes more prominent at larger rotation radii. Therefore, it can be concluded that increasing the rotation radius promotes the dissociation of bound states, and the dissociation perpendicular to the direction of rotational angular velocity is more significant than that parallel to it.
\begin{figure}[H]
	\centering
	\includegraphics[width=0.75\textwidth]{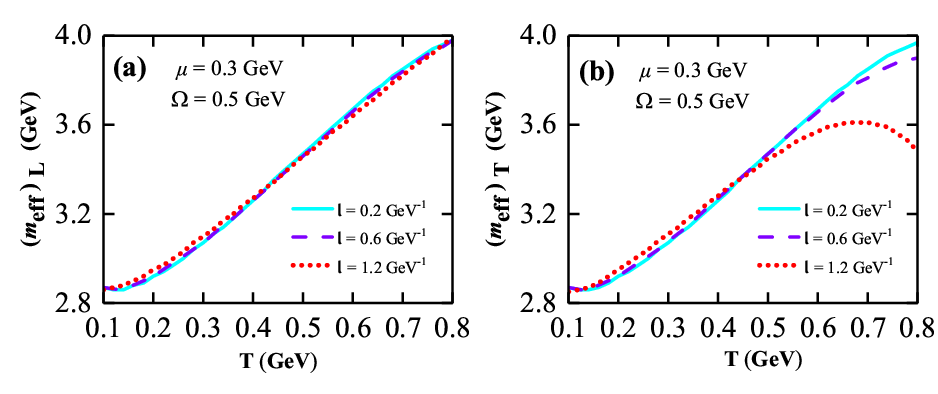}
	\caption{\label{fig3} The dependence of the effective mass on temperature at $\mu$ = 0.3 GeV
		and $\Omega$ = 0.5 GeV under different rotation radius.}
\end{figure}

Figure 3 shows the dependence of the effective mass of the peak position of the $ J / \psi$ spectral function on temperature at different rotation radii for $\mu$ = 0.3 GeV and $\Omega$ = 0.5 GeV. The first interesting finding is that the effective mass shows a slightly downward trend at lower temperatures, our explanation for this phenomenon is as follows: at lower temperatures, quarks are usually in a lower energy state. As the temperature increases, thermal excitation causes quarks to excite to higher energy states, resulting in a rearrangement of internal energy levels. Thermal excitation can reduce the effective potential of bound states, resulting in a decrease in the effective mass of mesons.

At higher temperatures, the influence of temperature on the system is more significant. The increase in temperature leads to an increase in the distance between quark-antiquark pairs, thereby enhancing the interaction between quark-antiquark and the medium. The second interesting finding is that in both the transverse and longitudinal directions, in the lower temperature region $ ( T \leq $ 0.42 GeV), the effective mass slightly increases with the increase of the rotation radius, while in the higher temperature region $ ( T > $ 0.42 GeV), the effective mass slightly decreases with the increase of the rotation radius. However, for the transverse rotational direction, when the rotation radius reaches $ l $ = 1.2 GeV$^ { - 1 } $ , the effective mass has a maximum near $T$ = 0.65 GeV , and then decreases sharply with increasing temperature. Combining with Fig. 2, one finds that when the rotation radius reaches $ l $ = 1.2 GeV$^ { - 1 } $ , the peak of the transverse spectrum is very low, indicating $ J / \psi$ dissociation. This indicates that the dissociation radius of the transverse spectrum of the $ J / \psi$ is smaller than that of the longitudinal spectrum.

\subsection{The Result For $ \Upsilon ( 1 S )$ }
	\begin{figure}[H]
	\centering
	\includegraphics[width=0.7\textwidth]{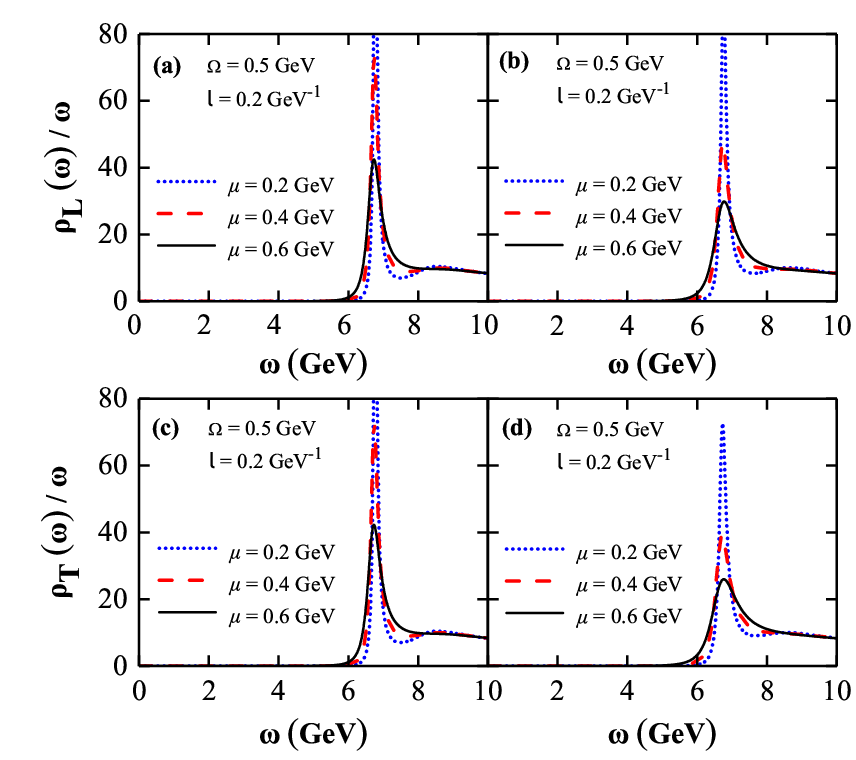}
	\caption{\label{fig4} Spectral functions of the $ \Upsilon ( 1 S )$ at $T = 0.265$ GeV, under different chemical potentials $\mu$ and rotational radius $l$. (a) and (b) are parallel to  the direction of rotation, while (c) and (d) are perpendicular to the direction of rotation. }
\end{figure}
\begin{figure}[H]
	\centering
	\includegraphics[width=0.6\textwidth]{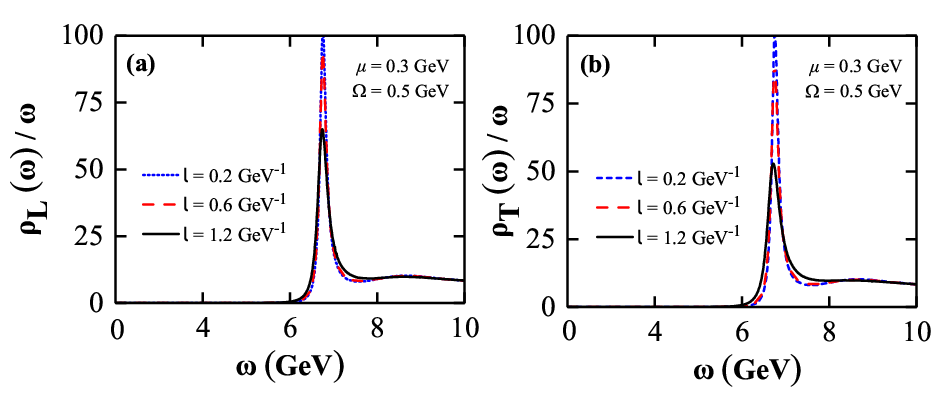}
	\caption{\label{fig5} Spectral functions of the $\Upsilon ( 1 S )$ for different rotation radius. At $\mu$ = 0.3 GeV and $\Omega$ = 0.5 GeV, it shows the spectral functions of the $ \Upsilon ( 1 S )$  states for different rotational radius at $T$ = 0.265 GeV. }
\end{figure}
\begin{figure}[H]
	\centering
	\includegraphics[width=0.6\textwidth]{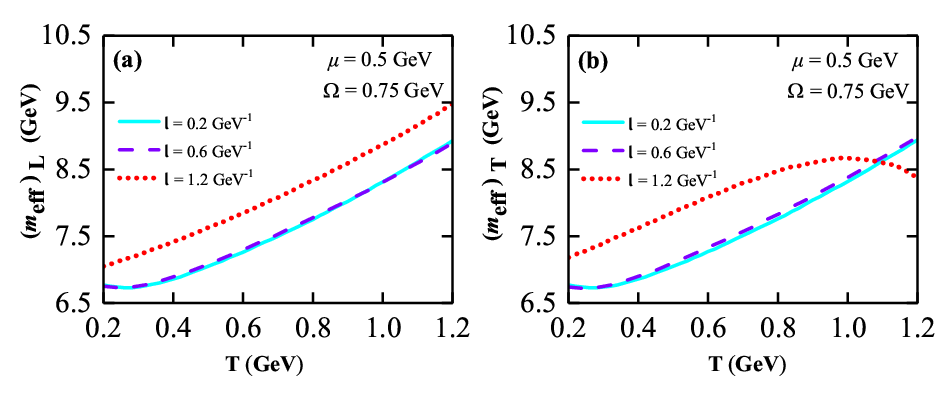}
	\caption{\label{fig6} The dependence of effective mass on temperature at $\mu$ = 0.5 GeV and $\Omega$ = 0.75 GeV,
		under different rotation radius.}
\end{figure}

Figures 4, and 5 show the spectral functions of longitudinal(L) and transverse(T) of the $\Upsilon ( 1 S )$ states under similar conditions. Compared with the previous results for the $ J / \psi$ state, it can be seen that the difference lies in the influence of the rotation radius on the peak position of the  $ \Upsilon ( 1 S )$ spectral function. These observations indicate that the increasing rotation radius in the rotating system can promote the dissociation process. 

As shown in Fig. 6, both the longitudinal and transverse effective masses increase with increasing temperature, which is consistent with the calculation results of the $ J / \psi$. When the rotation radius is small, the influence of the rotation radius on the relationship between effective mass and temperature is not significant in both longitudinal and transverse directions. However, as the rotation radius increases, the influence of the rotational radius on the relationship between effective mass and temperature becomes significant. The transverse effective mass varying with temperature shown in Fig. 6(b) exhibits a similar phenomenon to that of the transverse effective mass of the $ J / \psi$, reaching a maximum value when the temperature reaches a certain value ($T$ = 1 GeV).

\section{SUMMARY AND CONCLUSIONS}\label{sec:05 summary}

In this paper, we focus on the impact of rotation on the spectral function of the $ J / \psi$ and the $ \Upsilon ( 1 S )$ states. The metric by cylindrical coordinates with rotation is introduced into the system to calculate the spectral functions of heavy vector mesons. By incorporating rotation medium into the holographic QCD, we investigate the influence of rotational effect on the properties of the $ J / \psi$ and the $ \Upsilon ( 1 S )$ . From the results of this study, it is clear that temperature, chemical potential, and rotational radius effects enhance the dissociation process of heavy vector mesons within the medium. This rotation-induced effect is more significant for heavy vector mesons in the transverse direction.
\begin{figure}[H]
	\centering
	\includegraphics[width=0.6\textwidth]{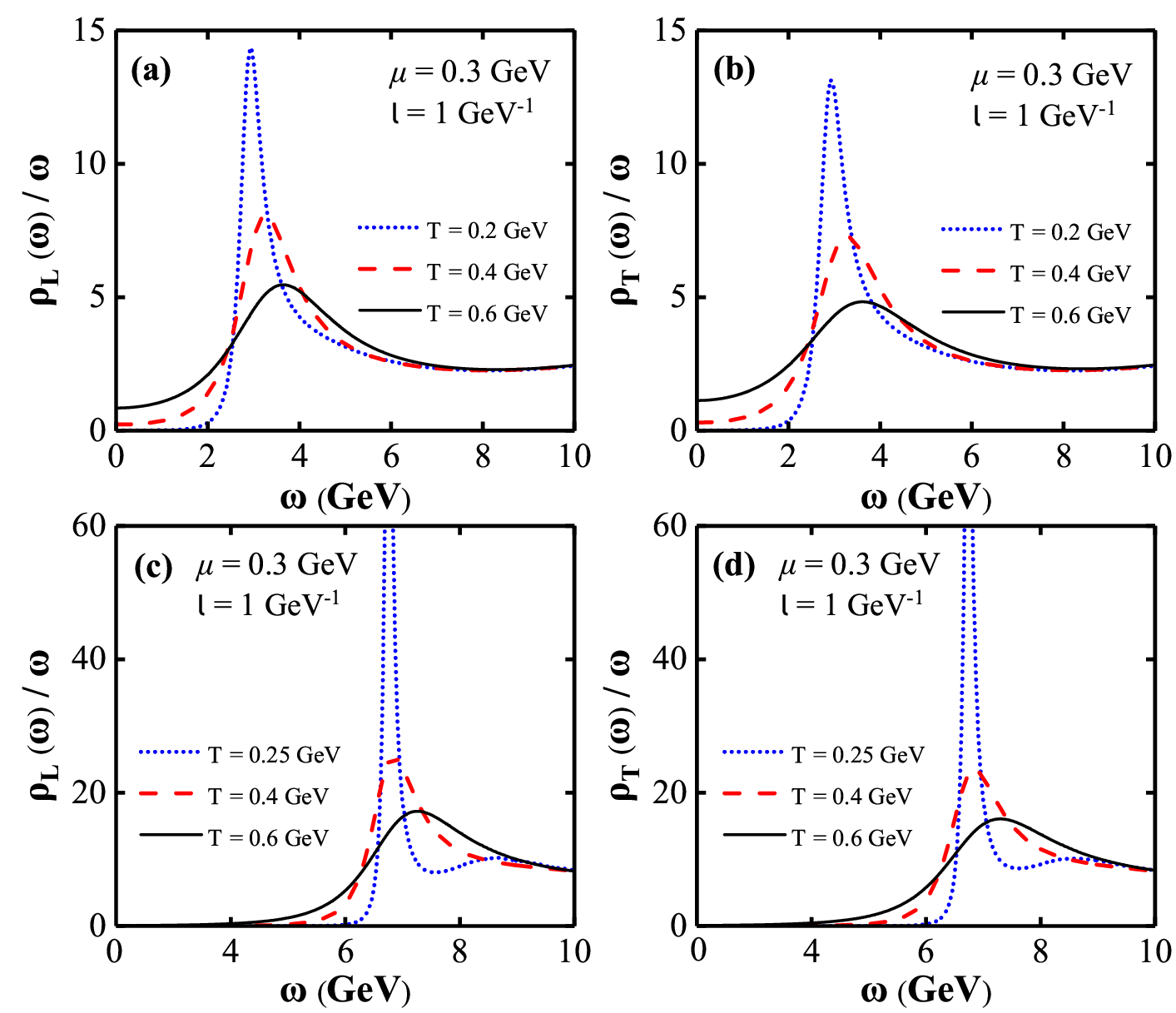}
	\caption{\label{fig7} The Spectral functions of the $ J / \psi$ and the $ \Upsilon ( 1 S )$ states for different temperatures
		$T$ at $\mu$ = 0.3 GeV , $\Omega$ = 0.5 GeV and $l$ = 1.0 GeV$^ { - 1 }$ .  (a) and (b) are for the $ J / \psi$ states, and (c) and (d) are for the $ \Upsilon ( 1 S )$ states; (a) and (c) are for the longitudinal direction, while (b) and (d) are for the transverse direction	
	}
\end{figure}

In this article, we mainly study the influence of the rotation radius on the spectral function. Here we need to add a discussion on the relationship between the spectral function and temperature, as shown in Fig.7. The spectral functions of the $ J / \psi$ and the $ \Upsilon ( 1 S )$ states with three different temperature values at $ \mu $ = 0.3 GeV, $\Omega$ = 0.5 GeV and $l$ = 1.0 GeV$^ { - 1 }$ are presented in Fig.7. This clearly demonstrates the impact of thermal effects on the dissociation process. It is found that as the temperature rises, the peak decreases and widens, and shifts to the right. This indicates that as the temperature increases, it promotes the dissociation of bound states and increases the decay width and effective mass.

The first holographic study on the influence of the radius of a homogeneous rotating system on the spectra functions for heavy vector mesons is established in the article.  As we are discussing the rotating system of QCD medium, the spectra functions characteristics should depend on the finite size of the rotating system. Due to the cylindrical symmetry of the rotating system, the rotation radius $l$ has become an important characteristic quantity of the rotating system. Studying the dependence of phase transition characteristics of the strongly interacting rotating matter on the radius of the rotating system is an important research topic. It is found that increasing the rotation radius promotes the dissociation of  the $ J / \psi$ and the $ \Upsilon ( 1 S )$ states, and the dissociation perpendicular to the direction of rotational angular velocity is more significant than that parallel to it at large rational radius.

	\section*{Acknowledgments}
	This work was supported by the National Natural Science Foundation of China (Grants No. 11875178, No. 11475068, No. 11747115).
	
	\section*{References}
	
	\nocite{*}
	\bibliography{ref}

\begin{thebibliography}{57}%
\makeatletter
\providecommand \@ifxundefined [1]{%
 \@ifx{#1\undefined}
}%
\providecommand \@ifnum [1]{%
 \ifnum #1\expandafter \@firstoftwo
 \else \expandafter \@secondoftwo
 \fi
}%
\providecommand \@ifx [1]{%
 \ifx #1\expandafter \@firstoftwo
 \else \expandafter \@secondoftwo
 \fi
}%
\providecommand \natexlab [1]{#1}%
\providecommand \enquote  [1]{``#1''}%
\providecommand \bibnamefont  [1]{#1}%
\providecommand \bibfnamefont [1]{#1}%
\providecommand \citenamefont [1]{#1}%
\providecommand \href@noop [0]{\@secondoftwo}%
\providecommand \href [0]{\begingroup \@sanitize@url \@href}%
\providecommand \@href[1]{\@@startlink{#1}\@@href}%
\providecommand \@@href[1]{\endgroup#1\@@endlink}%
\providecommand \@sanitize@url [0]{\catcode `\\12\catcode `\$12\catcode
  `\&12\catcode `\#12\catcode `\^12\catcode `\_12\catcode `\%12\relax}%
\providecommand \@@startlink[1]{}%
\providecommand \@@endlink[0]{}%
\providecommand \url  [0]{\begingroup\@sanitize@url \@url }%
\providecommand \@url [1]{\endgroup\@href {#1}{\urlprefix }}%
\providecommand \urlprefix  [0]{URL }%
\providecommand \Eprint [0]{\href }%
\providecommand \doibase [0]{http://dx.doi.org/}%
\providecommand \selectlanguage [0]{\@gobble}%
\providecommand \bibinfo  [0]{\@secondoftwo}%
\providecommand \bibfield  [0]{\@secondoftwo}%
\providecommand \translation [1]{[#1]}%
\providecommand \BibitemOpen [0]{}%
\providecommand \bibitemStop [0]{}%
\providecommand \bibitemNoStop [0]{.\EOS\space}%
\providecommand \EOS [0]{\spacefactor3000\relax}%
\providecommand \BibitemShut  [1]{\csname bibitem#1\endcsname}%
\let\auto@bib@innerbib\@empty
\bibitem [{\citenamefont {Adcox}\ \emph {et~al.}(2005)\citenamefont {Adcox}
  \emph {et~al.}}]{RN1}%
  \BibitemOpen
  \bibfield  {author} {\bibinfo {author} {\bibfnamefont {K.}~\bibnamefont
  {Adcox}} \emph {et~al.},\ }\href {\doibase 10.1016/j.nuclphysa.2005.03.086}
  {\bibfield  {journal} {\bibinfo  {journal} {Nucl. Phys. A}\ }\textbf
  {\bibinfo {volume} {757}},\ \bibinfo {pages} {184} (\bibinfo {year}
  {2005})}\BibitemShut {NoStop}%
\bibitem [{\citenamefont {Adams}\ \emph {et~al.}(2005)\citenamefont {Adams}
  \emph {et~al.}}]{RN2}%
  \BibitemOpen
  \bibfield  {author} {\bibinfo {author} {\bibfnamefont {J.}~\bibnamefont
  {Adams}} \emph {et~al.},\ }\href {\doibase 10.1016/j.nuclphysa.2005.03.085}
  {\bibfield  {journal} {\bibinfo  {journal} {Nucl. Phys. A}\ }\textbf
  {\bibinfo {volume} {757}},\ \bibinfo {pages} {102} (\bibinfo {year}
  {2005})}\BibitemShut {NoStop}%
\bibitem [{\citenamefont {Arsene}\ \emph {et~al.}(2005)\citenamefont {Arsene}
  \emph {et~al.}}]{RN3}%
  \BibitemOpen
  \bibfield  {author} {\bibinfo {author} {\bibfnamefont {I.}~\bibnamefont
  {Arsene}} \emph {et~al.},\ }\href {\doibase 10.1016/j.nuclphysa.2005.02.130}
  {\bibfield  {journal} {\bibinfo  {journal} {Nucl. Phys. A}\ }\textbf
  {\bibinfo {volume} {757}},\ \bibinfo {pages} {1} (\bibinfo {year}
  {2005})}\BibitemShut {NoStop}%
\bibitem [{\citenamefont {Matsui}\ and\ \citenamefont {Satz}(1986)}]{RN4}%
  \BibitemOpen
  \bibfield  {author} {\bibinfo {author} {\bibfnamefont {T.}~\bibnamefont
  {Matsui}}\ and\ \bibinfo {author} {\bibfnamefont {H.}~\bibnamefont {Satz}},\
  }\href {\doibase 10.1016/0370-2693(86)91404-8} {\bibfield  {journal}
  {\bibinfo  {journal} {Phys. Lett. B}\ }\textbf {\bibinfo {volume} {178}},\
  \bibinfo {pages} {416} (\bibinfo {year} {1986})}\BibitemShut {NoStop}%
\bibitem [{\citenamefont {Ma}(2019)}]{RN5}%
  \BibitemOpen
  \bibfield  {author} {\bibinfo {author} {\bibfnamefont {R.}~\bibnamefont
  {Ma}},\ }\href {\doibase 10.1016/j.nuclphysa.2018.07.010} {\bibfield
  {journal} {\bibinfo  {journal} {Nucl. Phys. A}\ }\textbf {\bibinfo {volume}
  {982}},\ \bibinfo {pages} {120} (\bibinfo {year} {2019})}\BibitemShut
  {NoStop}%
\bibitem [{\citenamefont {Karsch}(2005)}]{RN6}%
  \BibitemOpen
  \bibfield  {author} {\bibinfo {author} {\bibfnamefont {F.}~\bibnamefont
  {Karsch}},\ }\href {\doibase 10.1140/epjc/s2005-02192-2} {\bibfield
  {journal} {\bibinfo  {journal} {Eur. Phys. J. C}\ }\textbf {\bibinfo {volume}
  {43}},\ \bibinfo {pages} {35} (\bibinfo {year} {2005})}\BibitemShut {NoStop}%
\bibitem [{\citenamefont {Guo}\ \emph {et~al.}(2019)\citenamefont {Guo},
  \citenamefont {Shi}, \citenamefont {Feng},\ and\ \citenamefont {Liao}}]{RN7}%
  \BibitemOpen
  \bibfield  {author} {\bibinfo {author} {\bibfnamefont {Y.}~\bibnamefont
  {Guo}}, \bibinfo {author} {\bibfnamefont {S.}~\bibnamefont {Shi}}, \bibinfo
  {author} {\bibfnamefont {S.}~\bibnamefont {Feng}}, \ and\ \bibinfo {author}
  {\bibfnamefont {J.}~\bibnamefont {Liao}},\ }\href {\doibase
  10.1016/j.physletb.2019.134929} {\bibfield  {journal} {\bibinfo  {journal}
  {Phys. Lett. B}\ }\textbf {\bibinfo {volume} {798}},\ \bibinfo {pages}
  {134929} (\bibinfo {year} {2019})}\BibitemShut {NoStop}%
\bibitem [{\citenamefont {She}\ \emph {et~al.}(2018)\citenamefont {She},
  \citenamefont {Feng}, \citenamefont {Zhong},\ and\ \citenamefont
  {Yin}}]{RN8}%
  \BibitemOpen
  \bibfield  {author} {\bibinfo {author} {\bibfnamefont {D.}~\bibnamefont
  {She}}, \bibinfo {author} {\bibfnamefont {S.-Q.}\ \bibnamefont {Feng}},
  \bibinfo {author} {\bibfnamefont {Y.}~\bibnamefont {Zhong}}, \ and\ \bibinfo
  {author} {\bibfnamefont {Z.-B.}\ \bibnamefont {Yin}},\ }\href {\doibase
  10.1140/epja/i2018-12481-x} {\bibfield  {journal} {\bibinfo  {journal} {Eur.
  Phys. J. A}\ }\textbf {\bibinfo {volume} {54}},\ \bibinfo {pages} {48}
  (\bibinfo {year} {2018})}\BibitemShut {NoStop}%
\bibitem [{\citenamefont {Zhong}\ \emph {et~al.}(2014)\citenamefont {Zhong},
  \citenamefont {Yang}, \citenamefont {Cai},\ and\ \citenamefont {Feng}}]{RN9}%
  \BibitemOpen
  \bibfield  {author} {\bibinfo {author} {\bibfnamefont {Y.}~\bibnamefont
  {Zhong}}, \bibinfo {author} {\bibfnamefont {C.-B.}\ \bibnamefont {Yang}},
  \bibinfo {author} {\bibfnamefont {X.}~\bibnamefont {Cai}}, \ and\ \bibinfo
  {author} {\bibfnamefont {S.-Q.}\ \bibnamefont {Feng}},\ }\href {\doibase
  10.1155/2014/193039} {\bibfield  {journal} {\bibinfo  {journal} {Adv. High
  Energy Phys.}\ }\textbf {\bibinfo {volume} {2014}},\ \bibinfo {pages}
  {193039} (\bibinfo {year} {2014})}\BibitemShut {NoStop}%
\bibitem [{\citenamefont {Bzdak}\ and\ \citenamefont {Skokov}(2012)}]{RN10}%
  \BibitemOpen
  \bibfield  {author} {\bibinfo {author} {\bibfnamefont {A.}~\bibnamefont
  {Bzdak}}\ and\ \bibinfo {author} {\bibfnamefont {V.}~\bibnamefont {Skokov}},\
  }\href {\doibase 10.1016/j.physletb.2012.02.065} {\bibfield  {journal}
  {\bibinfo  {journal} {Phys. Lett. B}\ }\textbf {\bibinfo {volume} {710}},\
  \bibinfo {pages} {171} (\bibinfo {year} {2012})}\BibitemShut {NoStop}%
\bibitem [{\citenamefont {Deng}\ and\ \citenamefont {Huang}(2012)}]{RN11}%
  \BibitemOpen
  \bibfield  {author} {\bibinfo {author} {\bibfnamefont {W.-T.}\ \bibnamefont
  {Deng}}\ and\ \bibinfo {author} {\bibfnamefont {X.-G.}\ \bibnamefont
  {Huang}},\ }\href {\doibase 10.1103/PhysRevC.85.044907} {\bibfield  {journal}
  {\bibinfo  {journal} {Phys. Rev. C}\ }\textbf {\bibinfo {volume} {85}},\
  \bibinfo {pages} {044907} (\bibinfo {year} {2012})}\BibitemShut {NoStop}%
\bibitem [{\citenamefont {Wang}\ and\ \citenamefont {Feng}(2024)}]{RN12}%
  \BibitemOpen
  \bibfield  {author} {\bibinfo {author} {\bibfnamefont {J.-H.}\ \bibnamefont
  {Wang}}\ and\ \bibinfo {author} {\bibfnamefont {S.-Q.}\ \bibnamefont
  {Feng}},\ }\href {\doibase 10.1103/PhysRevD.109.066019} {\bibfield  {journal}
  {\bibinfo  {journal} {Phys. Rev. D}\ }\textbf {\bibinfo {volume} {109}},\
  \bibinfo {pages} {066019} (\bibinfo {year} {2024})}\BibitemShut {NoStop}%
\bibitem [{\citenamefont {Braga}\ \emph {et~al.}(2022)\citenamefont {Braga},
  \citenamefont {Faulhaber},\ and\ \citenamefont {Junqueira}}]{RN13}%
  \BibitemOpen
  \bibfield  {author} {\bibinfo {author} {\bibfnamefont {N.~R.~F.}\
  \bibnamefont {Braga}}, \bibinfo {author} {\bibfnamefont {L.~F.}\ \bibnamefont
  {Faulhaber}}, \ and\ \bibinfo {author} {\bibfnamefont {O.~C.}\ \bibnamefont
  {Junqueira}},\ }\href {\doibase 10.1103/PhysRevD.105.106003} {\bibfield
  {journal} {\bibinfo  {journal} {Phys. Rev. D}\ }\textbf {\bibinfo {volume}
  {105}},\ \bibinfo {pages} {106003} (\bibinfo {year} {2022})}\BibitemShut
  {NoStop}%
\bibitem [{\citenamefont {Bravo~Gaete}\ \emph {et~al.}(2017)\citenamefont
  {Bravo~Gaete}, \citenamefont {Guajardo},\ and\ \citenamefont
  {Hassaine}}]{RN14}%
  \BibitemOpen
  \bibfield  {author} {\bibinfo {author} {\bibfnamefont {M.}~\bibnamefont
  {Bravo~Gaete}}, \bibinfo {author} {\bibfnamefont {L.}~\bibnamefont
  {Guajardo}}, \ and\ \bibinfo {author} {\bibfnamefont {M.}~\bibnamefont
  {Hassaine}},\ }\href {\doibase 10.1007/JHEP04(2017)092} {\bibfield  {journal}
  {\bibinfo  {journal} {JHEP}\ }\textbf {\bibinfo {volume} {04}},\ \bibinfo
  {pages} {092} (\bibinfo {year} {2017})}\BibitemShut {NoStop}%
\bibitem [{\citenamefont {Erices}\ and\ \citenamefont {Martinez}(2018)}]{RN15}%
  \BibitemOpen
  \bibfield  {author} {\bibinfo {author} {\bibfnamefont {C.}~\bibnamefont
  {Erices}}\ and\ \bibinfo {author} {\bibfnamefont {C.}~\bibnamefont
  {Martinez}},\ }\href {\doibase 10.1103/PhysRevD.97.024034} {\bibfield
  {journal} {\bibinfo  {journal} {Phys. Rev. D}\ }\textbf {\bibinfo {volume}
  {97}},\ \bibinfo {pages} {024034} (\bibinfo {year} {2018})}\BibitemShut
  {NoStop}%
\bibitem [{\citenamefont {Braga}\ and\ \citenamefont {Ferreira}(2023)}]{RN16}%
  \BibitemOpen
  \bibfield  {author} {\bibinfo {author} {\bibfnamefont {N.~R.~F.}\
  \bibnamefont {Braga}}\ and\ \bibinfo {author} {\bibfnamefont {Y.~F.}\
  \bibnamefont {Ferreira}},\ }\href {\doibase 10.1103/PhysRevD.108.094017}
  {\bibfield  {journal} {\bibinfo  {journal} {Phys. Rev. D}\ }\textbf {\bibinfo
  {volume} {108}},\ \bibinfo {pages} {094017} (\bibinfo {year}
  {2023})}\BibitemShut {NoStop}%
\bibitem [{\citenamefont {Zhao}\ \emph {et~al.}(2023)\citenamefont {Zhao},
  \citenamefont {He}, \citenamefont {Hou}, \citenamefont {Li},\ and\
  \citenamefont {Li}}]{RN17}%
  \BibitemOpen
  \bibfield  {author} {\bibinfo {author} {\bibfnamefont {Y.-Q.}\ \bibnamefont
  {Zhao}}, \bibinfo {author} {\bibfnamefont {S.}~\bibnamefont {He}}, \bibinfo
  {author} {\bibfnamefont {D.}~\bibnamefont {Hou}}, \bibinfo {author}
  {\bibfnamefont {L.}~\bibnamefont {Li}}, \ and\ \bibinfo {author}
  {\bibfnamefont {Z.}~\bibnamefont {Li}},\ }\href {\doibase
  10.1007/JHEP04(2023)115} {\bibfield  {journal} {\bibinfo  {journal} {JHEP}\
  }\textbf {\bibinfo {volume} {04}},\ \bibinfo {pages} {115} (\bibinfo {year}
  {2023})}\BibitemShut {NoStop}%
\bibitem [{\citenamefont {Sun}\ and\ \citenamefont {Huang}(2022)}]{RN18}%
  \BibitemOpen
  \bibfield  {author} {\bibinfo {author} {\bibfnamefont {F.}~\bibnamefont
  {Sun}}\ and\ \bibinfo {author} {\bibfnamefont {A.}~\bibnamefont {Huang}},\
  }\href {\doibase 10.1103/PhysRevD.106.076007} {\bibfield  {journal} {\bibinfo
   {journal} {Phys. Rev. D}\ }\textbf {\bibinfo {volume} {106}},\ \bibinfo
  {pages} {076007} (\bibinfo {year} {2022})}\BibitemShut {NoStop}%
\bibitem [{\citenamefont {Jiang}\ and\ \citenamefont
  {Liao}(2016{\natexlab{a}})}]{RN19}%
  \BibitemOpen
  \bibfield  {author} {\bibinfo {author} {\bibfnamefont {Y.}~\bibnamefont
  {Jiang}}\ and\ \bibinfo {author} {\bibfnamefont {J.}~\bibnamefont {Liao}},\
  }\href {\doibase 10.1103/PhysRevLett.117.192302} {\bibfield  {journal}
  {\bibinfo  {journal} {Phys. Rev. Lett.}\ }\textbf {\bibinfo {volume} {117}},\
  \bibinfo {pages} {192302} (\bibinfo {year} {2016}{\natexlab{a}})}\BibitemShut
  {NoStop}%
\bibitem [{\citenamefont {Qiu}\ and\ \citenamefont {Feng}(2023)}]{RN20}%
  \BibitemOpen
  \bibfield  {author} {\bibinfo {author} {\bibfnamefont {Y.-W.}\ \bibnamefont
  {Qiu}}\ and\ \bibinfo {author} {\bibfnamefont {S.-Q.}\ \bibnamefont {Feng}},\
  }\href {\doibase 10.1103/PhysRevD.107.076004} {\bibfield  {journal} {\bibinfo
   {journal} {Phys. Rev. D}\ }\textbf {\bibinfo {volume} {107}},\ \bibinfo
  {pages} {076004} (\bibinfo {year} {2023})}\BibitemShut {NoStop}%
\bibitem [{\citenamefont {Qiu}\ \emph {et~al.}(2023)\citenamefont {Qiu},
  \citenamefont {Feng},\ and\ \citenamefont {Zhu}}]{RN21}%
  \BibitemOpen
  \bibfield  {author} {\bibinfo {author} {\bibfnamefont {Y.-W.}\ \bibnamefont
  {Qiu}}, \bibinfo {author} {\bibfnamefont {S.-Q.}\ \bibnamefont {Feng}}, \
  and\ \bibinfo {author} {\bibfnamefont {X.-Q.}\ \bibnamefont {Zhu}},\ }\href
  {\doibase 10.1103/PhysRevD.108.116022} {\bibfield  {journal} {\bibinfo
  {journal} {Phys. Rev. D}\ }\textbf {\bibinfo {volume} {108}},\ \bibinfo
  {pages} {116022} (\bibinfo {year} {2023})}\BibitemShut {NoStop}%
\bibitem [{\citenamefont {Bao}\ and\ \citenamefont {Feng}(2024)}]{RN22}%
  \BibitemOpen
  \bibfield  {author} {\bibinfo {author} {\bibfnamefont {Y.-R.}\ \bibnamefont
  {Bao}}\ and\ \bibinfo {author} {\bibfnamefont {S.-Q.}\ \bibnamefont {Feng}},\
  }\href {\doibase 10.1103/PhysRevD.109.096033} {\bibfield  {journal} {\bibinfo
   {journal} {Phys. Rev. D}\ }\textbf {\bibinfo {volume} {109}},\ \bibinfo
  {pages} {096033} (\bibinfo {year} {2024})}\BibitemShut {NoStop}%
\bibitem [{\citenamefont {Zhu}\ and\ \citenamefont {Feng}(2023)}]{RN23}%
  \BibitemOpen
  \bibfield  {author} {\bibinfo {author} {\bibfnamefont {X.}~\bibnamefont
  {Zhu}}\ and\ \bibinfo {author} {\bibfnamefont {S.-Q.}\ \bibnamefont {Feng}},\
  }\href {\doibase 10.1103/PhysRevD.107.016018} {\bibfield  {journal} {\bibinfo
   {journal} {Phys. Rev. D}\ }\textbf {\bibinfo {volume} {107}},\ \bibinfo
  {pages} {016018} (\bibinfo {year} {2023})}\BibitemShut {NoStop}%
\bibitem [{\citenamefont {Watts}\ \emph {et~al.}(2016)\citenamefont {Watts}
  \emph {et~al.}}]{RN24}%
  \BibitemOpen
  \bibfield  {author} {\bibinfo {author} {\bibfnamefont {A.~L.}\ \bibnamefont
  {Watts}} \emph {et~al.},\ }\href {\doibase 10.1103/RevModPhys.88.021001}
  {\bibfield  {journal} {\bibinfo  {journal} {Rev. Mod. Phys.}\ }\textbf
  {\bibinfo {volume} {88}},\ \bibinfo {pages} {021001} (\bibinfo {year}
  {2016})}\BibitemShut {NoStop}%
\bibitem [{\citenamefont {Grenier}\ and\ \citenamefont {Harding}(2015)}]{RN25}%
  \BibitemOpen
  \bibfield  {author} {\bibinfo {author} {\bibfnamefont {I.~A.}\ \bibnamefont
  {Grenier}}\ and\ \bibinfo {author} {\bibfnamefont {A.~K.}\ \bibnamefont
  {Harding}},\ }\href {\doibase 10.1016/j.crhy.2015.08.013} {\bibfield
  {journal} {\bibinfo  {journal} {Comptes Rendus Physique}\ }\textbf {\bibinfo
  {volume} {16}},\ \bibinfo {pages} {641} (\bibinfo {year} {2015})}\BibitemShut
  {NoStop}%
\bibitem [{\citenamefont {Liang}\ and\ \citenamefont {Wang}(2005)}]{RN26}%
  \BibitemOpen
  \bibfield  {author} {\bibinfo {author} {\bibfnamefont {Z.-T.}\ \bibnamefont
  {Liang}}\ and\ \bibinfo {author} {\bibfnamefont {X.-N.}\ \bibnamefont
  {Wang}},\ }\href {\doibase 10.1103/PhysRevLett.94.102301} {\bibfield
  {journal} {\bibinfo  {journal} {Phys. Rev. Lett.}\ }\textbf {\bibinfo
  {volume} {94}},\ \bibinfo {pages} {102301} (\bibinfo {year}
  {2005})}\BibitemShut {NoStop}%
\bibitem [{\citenamefont {Huang}\ \emph {et~al.}(2011)\citenamefont {Huang},
  \citenamefont {Huovinen},\ and\ \citenamefont {Wang}}]{RN27}%
  \BibitemOpen
  \bibfield  {author} {\bibinfo {author} {\bibfnamefont {X.-G.}\ \bibnamefont
  {Huang}}, \bibinfo {author} {\bibfnamefont {P.}~\bibnamefont {Huovinen}}, \
  and\ \bibinfo {author} {\bibfnamefont {X.-N.}\ \bibnamefont {Wang}},\ }\href
  {\doibase 10.1103/PhysRevC.84.054910} {\bibfield  {journal} {\bibinfo
  {journal} {Phys. Rev. C}\ }\textbf {\bibinfo {volume} {84}},\ \bibinfo
  {pages} {054910} (\bibinfo {year} {2011})}\BibitemShut {NoStop}%
\bibitem [{\citenamefont {Becattini}\ \emph {et~al.}(2008)\citenamefont
  {Becattini}, \citenamefont {Piccinini},\ and\ \citenamefont {Rizzo}}]{RN28}%
  \BibitemOpen
  \bibfield  {author} {\bibinfo {author} {\bibfnamefont {F.}~\bibnamefont
  {Becattini}}, \bibinfo {author} {\bibfnamefont {F.}~\bibnamefont
  {Piccinini}}, \ and\ \bibinfo {author} {\bibfnamefont {J.}~\bibnamefont
  {Rizzo}},\ }\href {\doibase 10.1103/PhysRevC.77.024906} {\bibfield  {journal}
  {\bibinfo  {journal} {Phys. Rev. C}\ }\textbf {\bibinfo {volume} {77}},\
  \bibinfo {pages} {024906} (\bibinfo {year} {2008})}\BibitemShut {NoStop}%
\bibitem [{\citenamefont {Jiang}\ and\ \citenamefont
  {Liao}(2016{\natexlab{b}})}]{RN29}%
  \BibitemOpen
  \bibfield  {author} {\bibinfo {author} {\bibfnamefont {Y.}~\bibnamefont
  {Jiang}}\ and\ \bibinfo {author} {\bibfnamefont {J.}~\bibnamefont {Liao}},\
  }\href {\doibase 10.1103/PhysRevLett.117.192302} {\bibfield  {journal}
  {\bibinfo  {journal} {Phys. Rev. Lett.}\ }\textbf {\bibinfo {volume} {117}},\
  \bibinfo {pages} {192302} (\bibinfo {year} {2016}{\natexlab{b}})}\BibitemShut
  {NoStop}%
\bibitem [{\citenamefont {Pang}\ \emph {et~al.}(2016)\citenamefont {Pang},
  \citenamefont {Petersen}, \citenamefont {Wang},\ and\ \citenamefont
  {Wang}}]{RN30}%
  \BibitemOpen
  \bibfield  {author} {\bibinfo {author} {\bibfnamefont {L.-G.}\ \bibnamefont
  {Pang}}, \bibinfo {author} {\bibfnamefont {H.}~\bibnamefont {Petersen}},
  \bibinfo {author} {\bibfnamefont {Q.}~\bibnamefont {Wang}}, \ and\ \bibinfo
  {author} {\bibfnamefont {X.-N.}\ \bibnamefont {Wang}},\ }\href {\doibase
  10.1103/PhysRevLett.117.192301} {\bibfield  {journal} {\bibinfo  {journal}
  {Phys. Rev. Lett.}\ }\textbf {\bibinfo {volume} {117}},\ \bibinfo {pages}
  {192301} (\bibinfo {year} {2016})}\BibitemShut {NoStop}%
\bibitem [{\citenamefont {Csernai}\ \emph {et~al.}(2013)\citenamefont
  {Csernai}, \citenamefont {Magas},\ and\ \citenamefont {Wang}}]{RN31}%
  \BibitemOpen
  \bibfield  {author} {\bibinfo {author} {\bibfnamefont {L.~P.}\ \bibnamefont
  {Csernai}}, \bibinfo {author} {\bibfnamefont {V.~K.}\ \bibnamefont {Magas}},
  \ and\ \bibinfo {author} {\bibfnamefont {D.~J.}\ \bibnamefont {Wang}},\
  }\href {\doibase 10.1103/PhysRevC.87.034906} {\bibfield  {journal} {\bibinfo
  {journal} {Phys. Rev. C}\ }\textbf {\bibinfo {volume} {87}},\ \bibinfo
  {pages} {034906} (\bibinfo {year} {2013})}\BibitemShut {NoStop}%
\bibitem [{\citenamefont {Deng}\ and\ \citenamefont {Huang}(2016)}]{RN32}%
  \BibitemOpen
  \bibfield  {author} {\bibinfo {author} {\bibfnamefont {W.-T.}\ \bibnamefont
  {Deng}}\ and\ \bibinfo {author} {\bibfnamefont {X.-G.}\ \bibnamefont
  {Huang}},\ }\href {\doibase 10.1103/PhysRevC.93.064907} {\bibfield  {journal}
  {\bibinfo  {journal} {Phys. Rev. C}\ }\textbf {\bibinfo {volume} {93}},\
  \bibinfo {pages} {064907} (\bibinfo {year} {2016})}\BibitemShut {NoStop}%
\bibitem [{\citenamefont {Yamamoto}\ and\ \citenamefont {Hirono}(2013)}]{RN33}%
  \BibitemOpen
  \bibfield  {author} {\bibinfo {author} {\bibfnamefont {A.}~\bibnamefont
  {Yamamoto}}\ and\ \bibinfo {author} {\bibfnamefont {Y.}~\bibnamefont
  {Hirono}},\ }\href {\doibase 10.1103/PhysRevLett.111.081601} {\bibfield
  {journal} {\bibinfo  {journal} {Phys. Rev. Lett.}\ }\textbf {\bibinfo
  {volume} {111}},\ \bibinfo {pages} {081601} (\bibinfo {year}
  {2013})}\BibitemShut {NoStop}%
\bibitem [{\citenamefont {Deng}\ and\ \citenamefont {Feng}(2022)}]{RN34}%
  \BibitemOpen
  \bibfield  {author} {\bibinfo {author} {\bibfnamefont {J.}~\bibnamefont
  {Deng}}\ and\ \bibinfo {author} {\bibfnamefont {S.-Q.}\ \bibnamefont
  {Feng}},\ }\href {\doibase 10.1103/PhysRevD.105.026015} {\bibfield  {journal}
  {\bibinfo  {journal} {Phys. Rev. D}\ }\textbf {\bibinfo {volume} {105}},\
  \bibinfo {pages} {026015} (\bibinfo {year} {2022})}\BibitemShut {NoStop}%
\bibitem [{\citenamefont {Chen}\ \emph {et~al.}(2018)\citenamefont {Chen},
  \citenamefont {Feng}, \citenamefont {Shi},\ and\ \citenamefont
  {Zhong}}]{RN35}%
  \BibitemOpen
  \bibfield  {author} {\bibinfo {author} {\bibfnamefont {X.}~\bibnamefont
  {Chen}}, \bibinfo {author} {\bibfnamefont {S.-Q.}\ \bibnamefont {Feng}},
  \bibinfo {author} {\bibfnamefont {Y.-F.}\ \bibnamefont {Shi}}, \ and\
  \bibinfo {author} {\bibfnamefont {Y.}~\bibnamefont {Zhong}},\ }\href
  {\doibase 10.1103/PhysRevD.97.066015} {\bibfield  {journal} {\bibinfo
  {journal} {Phys. Rev. D}\ }\textbf {\bibinfo {volume} {97}},\ \bibinfo
  {pages} {066015} (\bibinfo {year} {2018})}\BibitemShut {NoStop}%
\bibitem [{\citenamefont {Chen}\ \emph {et~al.}(2021)\citenamefont {Chen},
  \citenamefont {Zhang}, \citenamefont {Li}, \citenamefont {Hou},\ and\
  \citenamefont {Huang}}]{RN36}%
  \BibitemOpen
  \bibfield  {author} {\bibinfo {author} {\bibfnamefont {X.}~\bibnamefont
  {Chen}}, \bibinfo {author} {\bibfnamefont {L.}~\bibnamefont {Zhang}},
  \bibinfo {author} {\bibfnamefont {D.}~\bibnamefont {Li}}, \bibinfo {author}
  {\bibfnamefont {D.}~\bibnamefont {Hou}}, \ and\ \bibinfo {author}
  {\bibfnamefont {M.}~\bibnamefont {Huang}},\ }\href {\doibase
  10.1007/JHEP07(2021)132} {\bibfield  {journal} {\bibinfo  {journal} {JHEP}\
  }\textbf {\bibinfo {volume} {07}},\ \bibinfo {pages} {132} (\bibinfo {year}
  {2021})}\BibitemShut {NoStop}%
\bibitem [{\citenamefont {Finazzo}\ \emph {et~al.}(2015)\citenamefont
  {Finazzo}, \citenamefont {Rougemont}, \citenamefont {Marrochio},\ and\
  \citenamefont {Noronha}}]{RN37}%
  \BibitemOpen
  \bibfield  {author} {\bibinfo {author} {\bibfnamefont {S.~I.}\ \bibnamefont
  {Finazzo}}, \bibinfo {author} {\bibfnamefont {R.}~\bibnamefont {Rougemont}},
  \bibinfo {author} {\bibfnamefont {H.}~\bibnamefont {Marrochio}}, \ and\
  \bibinfo {author} {\bibfnamefont {J.}~\bibnamefont {Noronha}},\ }\href
  {\doibase 10.1007/JHEP02(2015)051} {\bibfield  {journal} {\bibinfo  {journal}
  {JHEP}\ }\textbf {\bibinfo {volume} {02}},\ \bibinfo {pages} {051} (\bibinfo
  {year} {2015})}\BibitemShut {NoStop}%
\bibitem [{\citenamefont {Li}\ \emph {et~al.}(2015)\citenamefont {Li},
  \citenamefont {He},\ and\ \citenamefont {Huang}}]{RN38}%
  \BibitemOpen
  \bibfield  {author} {\bibinfo {author} {\bibfnamefont {D.}~\bibnamefont
  {Li}}, \bibinfo {author} {\bibfnamefont {S.}~\bibnamefont {He}}, \ and\
  \bibinfo {author} {\bibfnamefont {M.}~\bibnamefont {Huang}},\ }\href
  {\doibase 10.1007/JHEP06(2015)046} {\bibfield  {journal} {\bibinfo  {journal}
  {JHEP}\ }\textbf {\bibinfo {volume} {06}},\ \bibinfo {pages} {046} (\bibinfo
  {year} {2015})}\BibitemShut {NoStop}%
\bibitem [{\citenamefont {Giataganas}\ \emph {et~al.}(2018)\citenamefont
  {Giataganas}, \citenamefont {G\"ursoy},\ and\ \citenamefont
  {Pedraza}}]{RN39}%
  \BibitemOpen
  \bibfield  {author} {\bibinfo {author} {\bibfnamefont {D.}~\bibnamefont
  {Giataganas}}, \bibinfo {author} {\bibfnamefont {U.}~\bibnamefont
  {G\"ursoy}}, \ and\ \bibinfo {author} {\bibfnamefont {J.~F.}\ \bibnamefont
  {Pedraza}},\ }\href {\doibase 10.1103/PhysRevLett.121.121601} {\bibfield
  {journal} {\bibinfo  {journal} {Phys. Rev. Lett.}\ }\textbf {\bibinfo
  {volume} {121}},\ \bibinfo {pages} {121601} (\bibinfo {year}
  {2018})}\BibitemShut {NoStop}%
\bibitem [{\citenamefont {Karch}\ \emph {et~al.}(2006)\citenamefont {Karch},
  \citenamefont {Katz}, \citenamefont {Son},\ and\ \citenamefont
  {Stephanov}}]{RN40}%
  \BibitemOpen
  \bibfield  {author} {\bibinfo {author} {\bibfnamefont {A.}~\bibnamefont
  {Karch}}, \bibinfo {author} {\bibfnamefont {E.}~\bibnamefont {Katz}},
  \bibinfo {author} {\bibfnamefont {D.~T.}\ \bibnamefont {Son}}, \ and\
  \bibinfo {author} {\bibfnamefont {M.~A.}\ \bibnamefont {Stephanov}},\ }\href
  {\doibase 10.1103/PhysRevD.74.015005} {\bibfield  {journal} {\bibinfo
  {journal} {Phys. Rev. D}\ }\textbf {\bibinfo {volume} {74}},\ \bibinfo
  {pages} {015005} (\bibinfo {year} {2006})}\BibitemShut {NoStop}%
\bibitem [{\citenamefont {Vega}\ and\ \citenamefont
  {Martin~Contreras}(2019)}]{RN41}%
  \BibitemOpen
  \bibfield  {author} {\bibinfo {author} {\bibfnamefont {A.}~\bibnamefont
  {Vega}}\ and\ \bibinfo {author} {\bibfnamefont {M.~A.}\ \bibnamefont
  {Martin~Contreras}},\ }\href {\doibase 10.1016/j.nuclphysb.2019.03.014}
  {\bibfield  {journal} {\bibinfo  {journal} {Nucl. Phys. B}\ }\textbf
  {\bibinfo {volume} {942}},\ \bibinfo {pages} {410} (\bibinfo {year}
  {2019})}\BibitemShut {NoStop}%
\bibitem [{\citenamefont {Martin~Contreras}\ \emph
  {et~al.}(2021{\natexlab{a}})\citenamefont {Martin~Contreras}, \citenamefont
  {Diles},\ and\ \citenamefont {Vega}}]{RN42}%
  \BibitemOpen
  \bibfield  {author} {\bibinfo {author} {\bibfnamefont {M.~A.}\ \bibnamefont
  {Martin~Contreras}}, \bibinfo {author} {\bibfnamefont {S.}~\bibnamefont
  {Diles}}, \ and\ \bibinfo {author} {\bibfnamefont {A.}~\bibnamefont {Vega}},\
  }\href {\doibase 10.1103/PhysRevD.103.086008} {\bibfield  {journal} {\bibinfo
   {journal} {Phys. Rev. D}\ }\textbf {\bibinfo {volume} {103}},\ \bibinfo
  {pages} {086008} (\bibinfo {year} {2021}{\natexlab{a}})}\BibitemShut
  {NoStop}%
\bibitem [{\citenamefont {Mamani}\ \emph {et~al.}(2022)\citenamefont {Mamani},
  \citenamefont {Hou},\ and\ \citenamefont {Braga}}]{RN43}%
  \BibitemOpen
  \bibfield  {author} {\bibinfo {author} {\bibfnamefont {L.~A.~H.}\
  \bibnamefont {Mamani}}, \bibinfo {author} {\bibfnamefont {D.}~\bibnamefont
  {Hou}}, \ and\ \bibinfo {author} {\bibfnamefont {N.~R.~F.}\ \bibnamefont
  {Braga}},\ }\href {\doibase 10.1103/PhysRevD.105.126020} {\bibfield
  {journal} {\bibinfo  {journal} {Phys. Rev. D}\ }\textbf {\bibinfo {volume}
  {105}},\ \bibinfo {pages} {126020} (\bibinfo {year} {2022})}\BibitemShut
  {NoStop}%
\bibitem [{\citenamefont {Fujita}\ \emph {et~al.}(2010)\citenamefont {Fujita},
  \citenamefont {Kikuchi}, \citenamefont {Fukushima}, \citenamefont {Misumi},\
  and\ \citenamefont {Murata}}]{RN44}%
  \BibitemOpen
  \bibfield  {author} {\bibinfo {author} {\bibfnamefont {M.}~\bibnamefont
  {Fujita}}, \bibinfo {author} {\bibfnamefont {T.}~\bibnamefont {Kikuchi}},
  \bibinfo {author} {\bibfnamefont {K.}~\bibnamefont {Fukushima}}, \bibinfo
  {author} {\bibfnamefont {T.}~\bibnamefont {Misumi}}, \ and\ \bibinfo {author}
  {\bibfnamefont {M.}~\bibnamefont {Murata}},\ }\href {\doibase
  10.1103/PhysRevD.81.065024} {\bibfield  {journal} {\bibinfo  {journal} {Phys.
  Rev. D}\ }\textbf {\bibinfo {volume} {81}},\ \bibinfo {pages} {065024}
  (\bibinfo {year} {2010})}\BibitemShut {NoStop}%
\bibitem [{\citenamefont {Satz}(2006)}]{RN45}%
  \BibitemOpen
  \bibfield  {author} {\bibinfo {author} {\bibfnamefont {H.}~\bibnamefont
  {Satz}},\ }\href {\doibase 10.1088/0954-3899/32/3/R01} {\bibfield  {journal}
  {\bibinfo  {journal} {J. Phys. G}\ }\textbf {\bibinfo {volume} {32}},\
  \bibinfo {pages} {R25} (\bibinfo {year} {2006})}\BibitemShut {NoStop}%
\bibitem [{\citenamefont {Braga}\ and\ \citenamefont
  {Ferreira}(2019{\natexlab{a}})}]{RN46}%
  \BibitemOpen
  \bibfield  {author} {\bibinfo {author} {\bibfnamefont {N.~R.~F.}\
  \bibnamefont {Braga}}\ and\ \bibinfo {author} {\bibfnamefont {L.~F.}\
  \bibnamefont {Ferreira}},\ }\href {\doibase 10.1007/JHEP01(2019)082}
  {\bibfield  {journal} {\bibinfo  {journal} {JHEP}\ }\textbf {\bibinfo
  {volume} {01}},\ \bibinfo {pages} {082} (\bibinfo {year}
  {2019}{\natexlab{a}})}\BibitemShut {NoStop}%
\bibitem [{\citenamefont {Braga}\ and\ \citenamefont
  {Ferreira}(2019{\natexlab{b}})}]{RN47}%
  \BibitemOpen
  \bibfield  {author} {\bibinfo {author} {\bibfnamefont {N.~R.~F.}\
  \bibnamefont {Braga}}\ and\ \bibinfo {author} {\bibfnamefont {L.~F.}\
  \bibnamefont {Ferreira}},\ }\href {\doibase 10.1016/j.physletb.2019.06.050}
  {\bibfield  {journal} {\bibinfo  {journal} {Phys. Lett. B}\ }\textbf
  {\bibinfo {volume} {795}},\ \bibinfo {pages} {462} (\bibinfo {year}
  {2019}{\natexlab{b}})}\BibitemShut {NoStop}%
\bibitem [{\citenamefont {Martin~Contreras}\ \emph
  {et~al.}(2021{\natexlab{b}})\citenamefont {Martin~Contreras}, \citenamefont
  {Diles},\ and\ \citenamefont {Vega}}]{RN48}%
  \BibitemOpen
  \bibfield  {author} {\bibinfo {author} {\bibfnamefont {M.~A.}\ \bibnamefont
  {Martin~Contreras}}, \bibinfo {author} {\bibfnamefont {S.}~\bibnamefont
  {Diles}}, \ and\ \bibinfo {author} {\bibfnamefont {A.}~\bibnamefont {Vega}},\
  }\href {\doibase 10.1103/PhysRevD.103.086008} {\bibfield  {journal} {\bibinfo
   {journal} {Phys. Rev. D}\ }\textbf {\bibinfo {volume} {103}},\ \bibinfo
  {pages} {086008} (\bibinfo {year} {2021}{\natexlab{b}})}\BibitemShut
  {NoStop}%
\bibitem [{\citenamefont {Z\"ollner}\ and\ \citenamefont
  {K\"ampfer}(2021)}]{RN49}%
  \BibitemOpen
  \bibfield  {author} {\bibinfo {author} {\bibfnamefont {R.}~\bibnamefont
  {Z\"ollner}}\ and\ \bibinfo {author} {\bibfnamefont {B.}~\bibnamefont
  {K\"ampfer}},\ }\href {\doibase 10.1103/PhysRevD.104.106005} {\bibfield
  {journal} {\bibinfo  {journal} {Phys. Rev. D}\ }\textbf {\bibinfo {volume}
  {104}},\ \bibinfo {pages} {106005} (\bibinfo {year} {2021})}\BibitemShut
  {NoStop}%
\bibitem [{\citenamefont {Chernodub}(2021)}]{RN50}%
  \BibitemOpen
  \bibfield  {author} {\bibinfo {author} {\bibfnamefont {M.~N.}\ \bibnamefont
  {Chernodub}},\ }\href {\doibase 10.1103/PhysRevD.103.054027} {\bibfield
  {journal} {\bibinfo  {journal} {Phys. Rev. D}\ }\textbf {\bibinfo {volume}
  {103}},\ \bibinfo {pages} {054027} (\bibinfo {year} {2021})}\BibitemShut
  {NoStop}%
\bibitem [{\citenamefont {Zhu}\ \emph {et~al.}(2019)\citenamefont {Zhu},
  \citenamefont {Feng}, \citenamefont {Shi},\ and\ \citenamefont
  {Zhong}}]{RN51}%
  \BibitemOpen
  \bibfield  {author} {\bibinfo {author} {\bibfnamefont {Z.-R.}\ \bibnamefont
  {Zhu}}, \bibinfo {author} {\bibfnamefont {S.-Q.}\ \bibnamefont {Feng}},
  \bibinfo {author} {\bibfnamefont {Y.-F.}\ \bibnamefont {Shi}}, \ and\
  \bibinfo {author} {\bibfnamefont {Y.}~\bibnamefont {Zhong}},\ }\href
  {\doibase 10.1103/PhysRevD.99.126001} {\bibfield  {journal} {\bibinfo
  {journal} {Phys. Rev. D}\ }\textbf {\bibinfo {volume} {99}},\ \bibinfo
  {pages} {126001} (\bibinfo {year} {2019})}\BibitemShut {NoStop}%
\bibitem [{\citenamefont {Feng}\ \emph {et~al.}(2020)\citenamefont {Feng},
  \citenamefont {Zhao},\ and\ \citenamefont {Chen}}]{RN52}%
  \BibitemOpen
  \bibfield  {author} {\bibinfo {author} {\bibfnamefont {S.-Q.}\ \bibnamefont
  {Feng}}, \bibinfo {author} {\bibfnamefont {Y.-Q.}\ \bibnamefont {Zhao}}, \
  and\ \bibinfo {author} {\bibfnamefont {X.}~\bibnamefont {Chen}},\ }\href
  {\doibase 10.1103/PhysRevD.101.026023} {\bibfield  {journal} {\bibinfo
  {journal} {Phys. Rev. D}\ }\textbf {\bibinfo {volume} {101}},\ \bibinfo
  {pages} {026023} (\bibinfo {year} {2020})}\BibitemShut {NoStop}%
\bibitem [{\citenamefont {Zhao}\ and\ \citenamefont {Hou}(2023)}]{RN53}%
  \BibitemOpen
  \bibfield  {author} {\bibinfo {author} {\bibfnamefont {Y.-Q.}\ \bibnamefont
  {Zhao}}\ and\ \bibinfo {author} {\bibfnamefont {D.}~\bibnamefont {Hou}},\
  }\href {\doibase 10.1140/epjc/s10052-023-12250-y} {\bibfield  {journal}
  {\bibinfo  {journal} {Eur. Phys. J. C}\ }\textbf {\bibinfo {volume} {83}},\
  \bibinfo {pages} {1076} (\bibinfo {year} {2023})}\BibitemShut {NoStop}%
\bibitem [{\citenamefont {Zhou}\ \emph {et~al.}(2021)\citenamefont {Zhou},
  \citenamefont {Chen}, \citenamefont {Zhao},\ and\ \citenamefont
  {Ping}}]{RN54}%
  \BibitemOpen
  \bibfield  {author} {\bibinfo {author} {\bibfnamefont {J.}~\bibnamefont
  {Zhou}}, \bibinfo {author} {\bibfnamefont {X.}~\bibnamefont {Chen}}, \bibinfo
  {author} {\bibfnamefont {Y.-Q.}\ \bibnamefont {Zhao}}, \ and\ \bibinfo
  {author} {\bibfnamefont {J.}~\bibnamefont {Ping}},\ }\href {\doibase
  10.1103/PhysRevD.102.126029} {\bibfield  {journal} {\bibinfo  {journal}
  {Phys. Rev. D}\ }\textbf {\bibinfo {volume} {102}},\ \bibinfo {pages}
  {126029} (\bibinfo {year} {2021})}\BibitemShut {NoStop}%
\bibitem [{\citenamefont {Braga}\ and\ \citenamefont {Ferreira}(2018)}]{RN55}%
  \BibitemOpen
  \bibfield  {author} {\bibinfo {author} {\bibfnamefont {N.~R.~F.}\
  \bibnamefont {Braga}}\ and\ \bibinfo {author} {\bibfnamefont {L.~F.}\
  \bibnamefont {Ferreira}},\ }\href {\doibase 10.1016/j.physletb.2018.06.053}
  {\bibfield  {journal} {\bibinfo  {journal} {Phys. Lett. B}\ }\textbf
  {\bibinfo {volume} {783}},\ \bibinfo {pages} {186} (\bibinfo {year}
  {2018})}\BibitemShut {NoStop}%
\bibitem [{\citenamefont {Nadi}\ \emph {et~al.}(2019)\citenamefont {Nadi},
  \citenamefont {Mirza}, \citenamefont {Sherkatghanad},\ and\ \citenamefont
  {Mirzaiyan}}]{RN56}%
  \BibitemOpen
  \bibfield  {author} {\bibinfo {author} {\bibfnamefont {H.}~\bibnamefont
  {Nadi}}, \bibinfo {author} {\bibfnamefont {B.}~\bibnamefont {Mirza}},
  \bibinfo {author} {\bibfnamefont {Z.}~\bibnamefont {Sherkatghanad}}, \ and\
  \bibinfo {author} {\bibfnamefont {Z.}~\bibnamefont {Mirzaiyan}},\ }\href
  {\doibase 10.1016/j.nuclphysb.2019.114822} {\bibfield  {journal} {\bibinfo
  {journal} {Nucl. Phys. B}\ }\textbf {\bibinfo {volume} {949}},\ \bibinfo
  {pages} {114822} (\bibinfo {year} {2019})}\BibitemShut {NoStop}%
\bibitem [{\citenamefont {Iqbal}\ and\ \citenamefont {Liu}(2009)}]{RN57}%
  \BibitemOpen
  \bibfield  {author} {\bibinfo {author} {\bibfnamefont {N.}~\bibnamefont
  {Iqbal}}\ and\ \bibinfo {author} {\bibfnamefont {H.}~\bibnamefont {Liu}},\
  }\href {\doibase 10.1103/PhysRevD.79.025023} {\bibfield  {journal} {\bibinfo
  {journal} {Phys. Rev. D}\ }\textbf {\bibinfo {volume} {79}},\ \bibinfo
  {pages} {025023} (\bibinfo {year} {2009})}\BibitemShut {NoStop}%
\end{thebibliography}%
	
\end{document}